\documentstyle[aps,twocolumn,epsf]{revtex}

\begin{document}

\title{Non-universal corrections to the level curvature distribution
beyond random matrix theory.}
\date{\today}
\author{I.\  V.\  Yurkevich$^{1,3}$ and V.\ E.\ Kravtsov$^{1,2}$}
\address{$^1$ International Center for Theoretical Physics, P.O. Box
586, 34100 Trieste, Italy\\$^2$ 
Landau Institute for Theoretical Physics, Kosygina str. 2, 117940 Moscow,
Russia\\$^3$ Institute for Low Temp. Physics and Engineering, Lenin 
ave. 44, 310164 Kharkov, Ukraine.} 
\maketitle 
\begin{abstract}
The level curvature distribution function is studied beyond the random 
matrix theory for the case of T-breaking perturbations over the 
orthogonal ensemble. The leading correction to the shape of the level 
curvature distribution is calculated using the nonlinear sigma-model.
The sign of the correction depends on the presence or absence of the 
global gauge invariance  and is different for perturbations caused by the 
constant vector-potential and by the random magnetic 
field. Scaling arguments are discussed that indicate on the 
qualitative difference in the level statistics in the dirty metal phase for 
space dimensionalities $d<4$ and $d>4$.
\end{abstract}
\draft
Since the seminal work by Thouless [1] the sensitivity of spectrum
 $\{E_{n}\}$ of
disordered conductors to a small twist  of the boundary conditions
$\Psi(\bf{\rho},z=0)=e^{i\varphi}\Psi(\bf{\rho},z=L)$ is considered as
a powerful tool to probe the space structure of eigenfunctions
and distinguish between the extended and the localized states. 
More precisely, the quantity $K_{n}$ which is now referred to as the "level 
curvature", was introduced in Ref.[1] in order to describe this sensitivity 
quantitatively:
\begin{equation}
\label{K}
K_{n}=\Delta^{-1}\,\left.\frac{\partial^2 
E_{n}(\varphi)}{\partial\varphi^2}\right| _{\varphi=0},
\end{equation}
where the mean level spacing $\Delta=(\nu L^{d})^{-1}$ is related with the 
mean  density of states $\nu$ and the size of the $d$-dimensional sample
$L$.

In complex or disordered quantum systems, the quantity $K_{n}$ fluctuates 
over the ensemble of energy  
levels $\{E_{n} \}$ or, for a given level, over the ensemble of 
realizations of 
disorder. The typical width of the distribution of level curvatures
$P(K)$ is of the order of the dimensionless conductance $g=D/(L^2\Delta)$,
where $D$ is the diffusion coefficient [1]. Thus studying this 
distribution one can obtain an information about the transition from 
metal to insulator with increasing the disorder (Anderson transition).

In the recent works [2] it has been shown that the distribution of the 
fluctuating quantity $K_{n}$ is a particular example of  {\it parametric 
level statistics}, i.e. statistics of spectral responses of the system to 
any perturbations proportional to some parameter $\varphi$. 

A remarkable property of the parametric level statistics [2] similar  
to that of the usual level statistics [3,4], is that in a certain limit 
they are universal for all classically chaotic and disordered systems and 
can be described by the random matrix theory (RMT) of Wigner and Dyson [3,4].
For disordered systems considered here this limit coincides [5] with 
$g\rightarrow\infty$. For chaotic systems the same role is played [6] by the
ratio $g=\gamma_{1}/\Delta$, where $\gamma_{1}$ is the first non-zero 
mode in the spectrum of 
the Perron-Frobenius operator that describes the chaotic behavior
of the corresponding classical system. In particular, for the 
time-reversal-invariant
system without spin-dependent interactions (orthogonal ensemble) the 
distribution of level curvatures, Eq.(\ref{K}), was found [7,8] in this limit
to have the form:
\begin{equation}
\label{RMT}
P_{WD}(k)=\frac{1}{2\,(1+ k^2)^{\frac{3}{2}}},\;\;\;\; 
k=\frac{K}{\langle|K|\rangle}, 
\end{equation}
where  $\langle|K|\rangle_{RMT}=2g$ is the average modulus of the level 
curvature.  
Further study [9] showed that the form, Eq.(\ref{RMT}), remains the same
if weak localization is taken into account, only the dimensionless 
conductance in the expression for $\langle|K|\rangle$ is changed 
appropriately.

The form, Eq.(\ref{RMT}), is universal. It does not depend, 
e.g. on  
details of the system and the perturbation. The only important point is that
the system is T-invariant with $T^2=1$ and the perturbation breaks this 
invariance.
However, the T-breaking perturbations can be of two different types.
One of them (we call such perturbations "global") is 
represented by the 
twist of the boundary conditions or by the magnetic flux in the problem of 
persistent current [10]. In this case  wave functions should have a 
significant amplitude on both boundaries $z=0$ and $z=L$ in order
the energy spectrum $E_{n}$ be sensitive to the phase $\varphi$.
Apparently, it is not so for the localized states, for which the 
distribution function P(k) shrinks dramatically and is approximately of the
logarithmically-normal form [11-13].  

Another type of T-breaking perturbations is represented by the magnetic 
impurities or random magnetic fluxes. We call them "local perturbations",
since they probe  wave functions locally. Even for a localized state 
there should be a significant spectral response to changing the strength 
$\varphi$ of such a local perturbation. So, one would  expect neither any 
dramatic contraction nor the logarithmically-normal form of the distribution 
function $P(K)$ in this case.

Thus we conclude that there is no universality of the 
distribution function
$P(k)$ in the Anderson insulator. It is reasonable to assume
that the universality is broken already at small but finite values of the 
parameter $1/g$ when the wave functions are still extended but have a 
pronounced
structure. In the presence of such a structure, the system cannot 
be described any more by the basis-invariant ensemble of random matrices
which is the main assumption of the random-matrix theory of Wigner and 
Dyson [3,4]. In order to use the curvature distribution function as a 
tool to probe the space structure of wave functions one should go beyond 
the random matrix theory. 

In the present Letter we address the problem of  corrections 
to the {\it shape} of the distribution function $P(k)$.
A general approach to
calculate such corrections using the nonlinear supersymmetric sigma-model 
[5] has been suggested in [14] and applied to 
distributions of different quantities [9,14,15]. It is 
based on the perturbative consideration of the non-zero diffusion modes 
which are
integrated out to produce corrections to the zero-mode supersymmetric 
sigma-model [5]. The latter is then integrated exactly using the Efetov's 
parametrization [5].  

Before going into details of the calculations we would like to formulate the 
main result and its consequences. The main result is the correction
to the level curvature distribution function $P(k)$ for the case of 
T-breaking perturbations over the orthogonal ensemble:
\begin{equation}
\label{MR}
\delta P(k)= \frac{C(d)}{(\pi g)^2}\,\;\frac{2-11 k^2 + 
2k^4}{2\,(1+k^2)^{7/2}}, \;\;\;\;\; k=\frac{K}{\langle|K|\rangle}\ll g, 
\end{equation}
where
\begin{equation}
\label{C}
C(d)=\sum_{{\bf q}\neq 0}\frac{1}{({\bf q}^2)^{2}}\times\left\{
\begin{array}{ll}
\left(\frac{4}{d}-1 \right),  & {\rm case\ I\ }\\ 
-1, & {\rm case\ II\ }
\end{array}\right.
\end{equation}
Here ${\bf q}=\{q_{1},...q_{d}\}$, where $q_{i}=2\pi n_{i}$, 
($n_{i}=0,\pm 1, \pm 2...$) in the case of the periodic  
boundary conditions (for an unperturbed system) considered in this Letter.

A remarkable feature of Eqs.(\ref{MR})-(\ref{C}) for $d<4$ is that the {\it 
sign} of the 
correction is different for global (case I) and local (case II) T-breaking 
perturbations. This reflects the
principal difference between them which we discussed above. The positive 
sign of the correction at small $K$ for the case of the global 
perturbations has been observed in the recent numerical simulations [12] on 
the 3D Anderson model. It implies the tendency towards the formation of a 
sharp peak in $P(K)$ at $K=0$ which should be present in the insulator for 
the case of global perturbations and absent in the case of local ones.

Next we note that the sum in Eq.(\ref{C}) is convergent only
for $d<4$. For $d>4$ there is a problem of the upper momentum cut-off
that cannot be done accurately within the framework of the nonlinear
sigma-model approach.
It turns out that for global perturbations the coefficient $C(d)$ changes 
sign precisely
at $d=4$ where the difficulty with the divergent sum is first encountered.
This coincidence could hardly be accidental.
It may indicate on the qualitative difference in the structure of 
wave functions in disordered systems for $d<4$ and $d>4$, especially near 
the Anderson transition where the parameter $1/g^2$ is not very small.  
Meanwhile the problem of the Anderson 
transition in higher dimensions is becoming a physical issue because
of some models [16] that map the problem of disordered systems of {\it 
interacted} electrons onto the Anderson model in higher dimensions.

Now we outline the derivation of the announced results. 
The first step is to express the level curvature distribution function
$P(K)$ in terms of the two-level correlation function $R(\omega,\varphi)$.
This is done in the same way as in Ref.[17]:
\begin{equation} 
\label{R-P}
P(K)=
\lim_{\varphi\rightarrow 
0}\left[
\frac{\varphi^2}{2}\,R\left(\omega=\frac{\varphi^2}{2}\,K\Delta;\;\;
\varphi\right)
\right].
\end{equation}  

The second step is to represent the two-level correlation function in 
Eq.(\ref{R-P}) in terms of 
the supersymmetric nonlinear sigma-model which has been derived 
in Ref.[5] from the model of non-interacting electrons in a random potential.
Using the results of Ref.[5] we arrive at:
\begin{equation}
\label{S-M R}
P(K)=-\left.\frac{\varphi^2}{16}\;\Re 
\frac{\partial^2}{\partial j_{+}\partial j_{-}}\int 
{\cal D}Q \;e^{-F_{\varphi}[Q;\,K,j_{\pm}]}\,\right|_
{\varphi, j_{\pm}\rightarrow 0 }.
\end{equation}
In this equation 
$Q({\bf r})$ is the $8\times 8$ supermatrix field of a certain symmetry 
which obeys the constraint $Q^2 =1$,
and 
$F_{\varphi}[Q;\,K,j_{\pm}]=F[Q;\,K]+F[Q;\,j_{+}]+F[Q;\,j_{-}]$, 
where: 
\begin{equation}
\label{Fj}
F[Q;\,j_{\pm}]=j_{\pm}\int Str[P_{\pm}P_{B}Q]\;\frac{d{\bf r}}{L^d}
\end{equation}
Here $Str[A]=tr[A_{FF}]-tr[A_{BB}]$ stands for the supertrace, $P_{\pm}$
are the projectors onto the retarded (RR) or advanced (AA)  
sectors  and $P_{B}$ is the projector onto the boson-boson
sector of the superfield $Q$ (see Ref.[5] for details and definitions).
An equivalent representation has been used in Ref.[9].   
 
The main information about the underlying physics is contained in the action
$F[Q;\,K]$. For global 
perturbations it  takes the form: 
\begin{eqnarray}
\label{F}
F_{\varphi}[Q;\,K]&=&\frac{\pi}{8}gL^2 \int Str\left({\vec\nabla} Q
-\frac{i{\vec\varphi}}{L}\,[Q,\tau] 
\right)^2\;\frac{d{\bf r}}{L^d} + \\ \nonumber 
&+& i\frac{\pi}{8}\,K\varphi^2\;\int Str[\Lambda Q]\;\frac{d{\bf 
r}}{L^d}, 
\end{eqnarray}
where ${\vec\varphi}=(0,\varphi)$ is supposed to be directed along the 
$z$-axis;
$\Lambda=P_{+}-P_{-}$, and $\tau=\sigma_{z}\otimes P_{+}$ is proportional 
to the Pauli matrix $\sigma_{z}$ that breaks the symmetry in the 
quaternionic subspace of the superfield $Q$ and reflects  
the T-symmetry breaking.

For local T-breaking perturbations the linear in $\varphi$ term in 
Eq.(\ref{F}) is absent.\\
The representation, Eqs.(\ref{S-M R})-(\ref{F}), in terms of the field
$Q({\bf r})$ contains all the spatial diffusion modes 
$\gamma_{q}=(D/L^2){\bf q}^{2}$.
However, in doing the limit $\varphi\rightarrow 0$ in Eq.(\ref{S-M R})
the main role is played by the zero mode which corresponds to ${\bf q}=0$.
At $\varphi=0$ this mode does not cost any energy no matter how large are 
the components of the field $Q$ in the non-compact boson-boson sector [5]. 
It is
the arbitrary large amplitudes of the zero mode components of the field $Q$ 
that compensate the infinitesimal parameter $\varphi$ in Eqs.(\ref{S-M 
R})-(\ref{F}) and lead to a finite result for $P(K)$. Thus the
space independent zero mode $Q_{0}$ must be 
considered
non-perturbatively. In the limit $g\rightarrow\infty$ all the non-zero 
modes can be neglected [5], and one arrives [9] at the RMT result, 
Eq.(\ref{RMT}). For finite $1/g$ the non-zero modes should be also taken 
into account.  However, all the non-zero modes can be treated
perturbatively for $g\gg 1$ to lead to some corrections to the zero-mode 
action. In order to obtain these corrections we have to separate zero modes 
from all other modes and then integrate over 
all the non-zero modes using a certain perturbative scheme.  
This is done by means of the transformation [14]:
\begin{equation}
\label{tr}
Q({\bf r})=V_{0}^{-1}\tilde{Q}({\bf r})V_{0},\;\;\;\;\; 
Q_{0}=V_{0}^{-1}\Lambda V_{0}. 
\end{equation}
The field $\tilde{Q}({\bf r})$ that must also obey the constraint 
$\tilde{Q}^2=1$, is parametrized as follows [14]: 
\begin{equation}
\label{par}
\tilde{Q}=(1-W/2)\Lambda (1-W/2)^{-1},
\end{equation}
where the supermatrix $W({\bf r})$ does not contain the zero spatial mode. 

The transformation from $Q({\bf r})$ to new variables $Q_{0}, W_{{\bf q\neq 
0}}$ involves [15] a non-trivial Jacobian $J=e^{-F_{J}[W]}$. To the 
first non-vanishing order in $W$ it is given by [18]: 
\begin{equation} 
\label{J}
F_{J}[W]= -\frac{1}{8}\int Str[W^2]\frac{d{\bf r}}{L^d}.
\end{equation}

In order to integrate over $W$ perturbatively, one has to expand the action,
Eqs.(\ref{Fj}),(\ref{F}),(\ref{J}) in powers of $W$ separating the 
$V_{0}$-independent quadratic in $W$
part $F_{0}[W]$ and 
consider all the rest $W$-dependent terms as perturbations using the 
Wick's theorem. 

As a result of such calculations we have obtained the correction up to the 
$1/g^2$ order to the zero-mode
action $F[Q_{0};\,K,j_{\pm}]$. It is 
too lengthy to 
present it here and contains all  possible linear and bi-linear 
combinations of vertices encountered in $F[Q_{0};\,K,j_{\pm}]$ plus
4 new vertices. 

A remarkable feature of the Efetov's parametrization
[5]
that has not been mentioned so far is that in 
the limit of large "non-compact
angles" $\lambda_{1,2}\rightarrow\infty$, the new vertices 
reduce to the product of the "old" ones. Thus up to the leading order
of $\lambda_{1,2}^4$ we found:  
\begin{eqnarray}
\label{dec}
& &Str[(\Lambda Q_{0})^2]= - \left[Str
[\Lambda Q_{0}]\right]^2,\\ \nonumber
& &
Str[(\tau Q_{0})^4]= -\frac{1}{2}\left[Str[(\tau Q_{0})^2]\right]^2 \\ 
\nonumber 
& & Str[(\tau Q_{0})^2 \Lambda Q_{0}]= -\frac{1}{2}\,Str[(\tau 
Q_{0})^2]\;Str
[\Lambda Q_{0}],\\ \nonumber
& & Str[(\tau Q_{0})^2 P_{\pm}P_{B}Q_{0}]= -\frac{1}{2}\,Str[(\tau Q_{0})^2]
\;Str[P_{\pm}P_{B}Q_{0}]. 
\end{eqnarray}
This circumstance simplifies the solution dramatically, since it allows to
obtain all the corrections to $P(K)$ just by 
differentiating of the RMT result, Eq.(\ref{RMT}), over $g$ and
$K$. Since the $K$ and  $g$-dependences  of 
the RMT distribution function $\tilde{P}_{WD}(K)=(2g)^{-1}P_{WD}(K/2g)$ are 
closely related,
the result can be formulated in terms of the $g$-derivatives alone. It is 
convenient to introduce the parameter $\alpha=\ln g$. Then the 
corrections to the distribution function $P(K)$ can be represented in the 
form:
\begin{eqnarray}
\label{ff}
& &\delta P(K)=-\frac{1}{\pi g}\sum_{{\bf q}\neq 0}\frac{1}{{\bf 
q}^2}\left(\frac{\partial \tilde{P}_{WD}(K)}{\partial\alpha}\right)+
\\ \nonumber
&+& \frac{1}{(\pi g)^2}\sum_{{\bf q}\neq 0}\frac{1}{({\bf q}^2)^2} 
\left[\frac{9}{2}-\frac{16}{d}+\frac{36}{d(d+2)} \right] 
\frac{\partial \tilde{P}_{WD}(K)}{\partial\alpha}+\\ \nonumber
&+&\frac{1}{(\pi g)^2}\sum_{{\bf q}\neq 0}\frac{1}{({\bf 
q}^2)^2}
\left(\frac{4}{d}-1 
\right)\;\frac{\partial^2 \tilde{P}_{WD}(K)}{\partial\alpha^2}. 
\end{eqnarray} 
The result, Eq.(\ref{ff}), holds for the global perturbations.
For such perturbations $\nabla Q$ enters Eq.(\ref{F}) as a
"covariant derivative"  $\nabla Q-\frac{i\varphi}{L}[Q,\tau]$. As a result,
the linear in $\varphi$ "cross"-vertex
$\vec{\varphi}\;Str[Q\vec{\nabla} Q\tau]$ appears in the action.
It is this vertex that leads to terms
in Eq.(\ref{ff}) proportional to $1/d$ or $1/d(d+2)$. For local
perturbations such terms are absent, since the linear in $\varphi$
vertex does not appear at all [5].
 
The term with the second derivative in Eq.(\ref{ff}) contributes to the
correction to the {\it shape} of the distribution function $P(k)$
and leads to the announced result, Eq.(\ref{MR}).

The terms proportional to the first derivative can be absorbed into
the RMT distribution function $P_{WD}(k)$ by changing the 
average $\langle |K|\rangle$. The $1/g$ correction  to 
this value first found in Ref.[9] is nothing but the usual 
weak-localization correction
to the quantity $\langle|K|\rangle_{RMT}=2g$. Thus up to the 
first order in  the inverse "bare" conductance $1/g$, the quantity  
$\frac{1}{2}\langle|K|\rangle$ 
behaves exactly like the true dimensionless conductance $\bar{g}$. 

Moreover, it turns out that all the $1/g^2$ corrections to $P(K)$ 
proportional to
$(\sum \frac{1}{{\bf q}^2})^2$ cancel each other in the 
same way as in the dimensionless conductance $\bar{g}$ [20]. 
This is in a full
agreement with the one-parameter scaling [21] extended to describe the
distributions of mesoscopic fluctuations [19]. In the region where 
the one-parameter scaling works, the distribution functions of all quantities
(measured in proper units) should depend only on one additional parameter 
- the
average dimensionless conductance $\bar{g}$ that absorbs all microscopic 
parameters of the system and includes all quantum corrections
proportional to the powers of $g^{-1}\sum\frac{1}{{\bf q}^2}$. For the 
case of the level curvature distribution it means that $P(K)=P(K;\bar{g})$,
where $\bar{g}$ should obey the scaling equation [21] that does not allow 
[20] the term 
proportional to $g^{-2}(\sum\frac{1}{{\bf q}^2})^2$ in the perturbative 
expansion of $\bar{g}/g$ in powers of $1/g$.

>From this viewpoint the corrections to $P(K)$ 
proportional to  $g^{-2} \sum\frac{1}{({\bf q}^2)^2}$ should be considered
as the {\it main} terms in the expansion of $P(K;\bar{g})$ in powers of the 
{\it true} inverse dimensionless conductance $1/\bar{g}$ 
renormalized by the 
localization effects rather than the second-order terms of an expansion in 
powers of the "bare" value of $1/g$.
With due regard of these corrections the shape of the distribution 
function $P(k)$ deviates from the RMT result and the average 
$\langle|K|\rangle$ is no more proportional to the average dimensionless 
conductance:
\begin{equation}
\label{av}
\langle|K|\rangle=2\bar{g}\left[1+\frac{A(d)}{(\pi \bar{g})^2}\sum_{{\bf 
q}\neq 0}\frac{1}{({\bf q}^2)^2} \right],
\end{equation}
where $A(d)=9/2$ for the local perturbations and $A(d)=(9/2-16/d + 
36/d(d+2))$ for the global perturbations.

In conclusion we would like to discuss a consequence of the form of  
corrections to different level statistics for the applicability of the RMT 
results. The typical relative correction [see e.g.[14,22]] has the same 
form as
Eq.(\ref{av}) and is proportional to $\bar{g}^{-2}\sum \frac{1}{({\bf 
q}^2)^2}$.
Let us suppose that the 
system is close enough to 
the Anderson transition (from the metal side) but it is still not in the 
critical region, so that $\pi \bar{g}\gg 1$ but 
it is not very large. Then for $d<4$, where the sum in Eq.(\ref{av}) is 
convergent, the correction is still small, and the RMT can be considered 
as a good zero approximation. However, it is not the case for $d>4$
if $\pi \bar{g} < (L/r_{0})^{d/2-2}$, where $r_{0}^{-1}$ is the upper 
momentum cut-off. In the scaling picture of the Anderson transition 
[21] the dimensionless conductance $\bar{g}$ can be made arbitrary close to
its critical value $g_{c}\sim 1$ by the fine tuning of the 
elastic scattering
mean free path $l\approx l_{c}$  at a 
constant $L\gg l\sim r_{0}$.  Thus we conclude that the above 
condition is always 
possible to achieve for $d>4$ at some strength of disorder  $W_{c1}(L)$ 
which is less than the critical one $W_{c}$. For $W>W_{c1}$
the RMT results do not work even as a zero 
approximation for the level statistics.  

In summary, we have calculated the leading corrections to the shape of 
the level curvature distribution function  
$P(K)$ beyond the random matrix 
theory. For T-breaking perturbations over the orthogonal ensemble, the 
corrections are not universal and depend 
qualitatively on the presence or absence of the global gauge invariance. 
In both cases the average $\frac{1}{2}\langle |K|\rangle$ shows deviations 
from the average quantum dimensionless conductance $\bar{g}$. 
The corrections are in agreement with the one-parameter scaling. The 
consequence of the scaling hypothesis and the form of the corrections is 
a qualitative difference in level statistics  
for space dimensionalities above and below $d=4$. Below $d=4$ the RMT is
a good zero approximation for level statistics in the metal phase, while for
$d\ge 4$ there should exist a critical value of disorder $W_{c1} < W_{c}$ 
above which RMT fails to describe spectral correlations.  

{\bf Acknowledgements} We thank B.L.Altshuler, Chaitali Basu, C.M.Canali, 
V.I.Fal'ko, 
Y.V.Fyodorov, I.V.Lerner and A.D.Mirlin for stimulated discussions.  
Support from RFBR and INTAS grant RFBR/INTAS 95-675 (V.E.K.) 
is gratefully acknowledged.


\begin{references}
\bibitem{18}
There was a minor
mistake
in calculating the Jacobian in Ref.[15]. The correct answer is given by
Eq.(\ref{J})
\end{references}
\end{document}